\documentclass[a4paper,11pt]{article}
\usepackage{pos}
\usepackage{mathtools}
\usepackage{textgreek}

\title{Alignment of the CMS Tracker: Results from LHC Run 3}
 \ShortTitle{Alignment of the CMS Tracker: Results from LHC Run 3}

\author*[a,1]{Sandra Consuegra Rodríguez}

\affiliation[a]{Deutsches Elektronen-Synchrotron,\\
  Notkestraße 85, 22607 Hamburg, Germany}
  
\note{On behalf of the CMS Collaboration}  

\emailAdd{sandra.consuegra.rodriguez@desy.de}

\abstract{The strategies for and the performance of the CMS tracker alignment during the ongoing Run 3 data-taking period are described. The results of the very first tracker alignment for Run 3  \nolinebreak data reprocessing performed with cosmic rays and collision tracks recorded at the unprecedented center of mass energy of 13.6 TeV are presented. Also, the performance after deployment of a more granular automated alignment associated with the improvement of the alignment calibration already during data taking is discussed. Finally, the prospects for the tracker alignment calibration during the Run 3 data-taking period, in light of the gained operational experience, are discussed.
}

\FullConference{%
  The European Physical Society Conference on High Energy Physics (EPS-HEP2023)\\
  21-25 August, 2023\\
  Hamburg, Germany
}

\begin{document}
\maketitle

\section{CMS tracker detector} 
The CMS detector at the CERN LHC features an all-silicon inner tracking system with a total active silicon area of $\sim$200 m$^{2}$, a length of 5.8 m, and a diameter of 2.5 m, called the tracker \cite{citation1}. The CMS tracker is composed of a Pixel detector at its innermost part and a Strip detector. The Pixel detector consists of four barrel layers at radii between 2.9 cm and 16 cm and three endcap disks per side, while the Strip detector consists of 10 barrel layers at radii between 20 cm and 116 cm, and twelve endcap disks per side. The pixel detector in the Phase 1 configuration is composed of 1856 modules, 1184 modules corresponding to the barrel pixel detector (BPIX), and 672 modules to the forward pixel detector (FPIX), for a total of 124 million readout channels \cite{citation2}, while the Strip detector is composed of 15148 modules and divided into four inner barrel (TIB) layers, three inner endcap disks (TID) per side, six outer barrel (TOB) layers (surrounding both the TIB and the TID), and 9 endcap disks (TEC) per side, for a total of 9.3 million readout channels \cite{citation3}.

\section{Track-based alignment} 
The size of each pixel (100x150 $\mu$m) composing the pixel detector allows the determination of the trajectory of a particle passing through the active area of the detector with a precision of up to $\sim$10 $\mu$m. Therefore, the alignment calibration should reach an accuracy significantly better than the intrinsic silicon hit resolution (10-30 $\mu$m) \cite{citation4}, in order to meet the physics goals of the experiment. The determination (correction) of the position, orientation, and surface deformations that describe the surface properties of the sensors of the tracker is the task of the tracker alignment calibration. The alignment parameters $\textbf{p}$ are derived  by minimizing the following $\chi^2$ function:
\begin{equation}
\chi^{2}(\textbf{p},\textbf{q})=\sum_{j}^{\text{tracks}}  \;   \sum_{i}^{\text{measurements}}
\left(\frac{m_{ij}-f_{ij}(\textbf{p},\textbf{q}_{j})}{\sigma_{ij}}\right)^2,
\label{eq:1}
\end{equation}
where the track hit-residual ($r_{ij}=m_{ij}-f_{ij}$) is obtained subtracting the projection of the track prediction $f_{ij}$ from the measured hit position $m_{ij}$, and $\sigma_{ij}$ corresponds to the $m_{ij}$ uncertainty \cite{citation5}. \\
During data-taking, the bulk of the main physics dataset is stored on disk for up to 48 hours before reconstruction, while a subset ("Stream Express"), together with dedicated calibration streams that make use of a reduced data format ("AlCaReco") go through a fast-track reconstruction. This streamlined reconstructed data is used to compute updated calibration conditions while the bulk of the data is still on disk. In this way, once the full prompt reconstruction starts it can profit from the already updated calibration constants, with the consequent improvement in the quality of the prompt reconstruction and reduction of the need for offline data reprocessing. This workflow is called Prompt Calibration Loop (PCL) and the tracker alignment calibration is integrated into it. \\
The track-based PCL automated alignment calibration in place during Run 2 and the beginning of Run 3 known as "Low Granularity (LG)" PCL alignment is based on a total of 36 degrees of freedom, to account for the high-level structure (HLS) movements of the Pixel detector, with 6 structures in total: two half barrels and two half cylinders in each of the two endcaps. For each of these structures, run-by-run corrections for position and rotation are derived (6X6). In September 2022, the "High Granularity (HG)" PCL alignment was deployed for production after a commissioning period, moving on from HLS-based to Pixel ladder/panel-based alignment, therefore, from 36 parameters to $\sim$5k parameters. The HG-PCL alignment which makes use of a brand new veto logic to cope with the larger amount of parameters, better accommodates and accounts for radiation effects during data taking, replacing manual alignments after pixel calibration updates. A more comprehensive calibration (including Strip detector alignment, time dependence, surface deformations, and tuning of the alignment position errors) is run offline exploiting the full dataset statistics, and therefore, profiting from increased kinematic variety to provide the final set of conditions for data reprocessing.

\subsection{Alignment algorithms: a complementary approach}
Two independent algorithms are used by CMS for tracker alignment, MillePede-II and HipPy, allowing the cross-check of the obtained results. MillePede-II, the algorithm of choice for the automated alignment and also used for offline calibration, performs a global fit including all correlations of global alignment parameters and local track parameters, while HipPy determines the alignment of individual sensors by minimizing a local $\chi^2$ function, therefore, with no large matrix inversions involved but with multiple iterations to account for the correlations between sensors  \cite{citation6}. 

\section{Performance in Run 3}
The alignment constants are determined in time intervals (IOVs) short enough to capture the motion of the substructures, but long enough to include sufficient data to obtain a precise calibration.

\subsection{Vertexing}
\begin{figure}[!ht]
    \centering
       \includegraphics[width=1\textwidth]{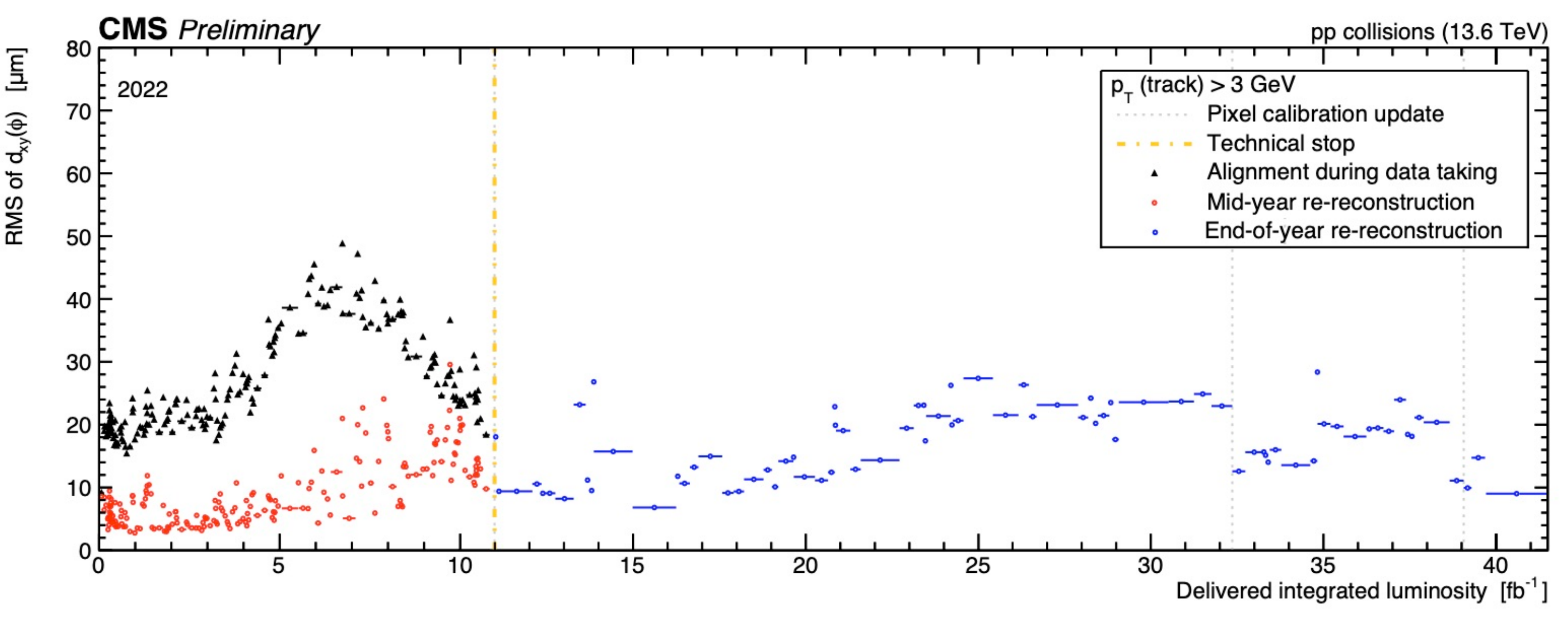}
       \caption{Mean track-vertex impact parameter RMS as function of delivered integrated luminosity \cite{citation7}.    
       }
       \label{fig1}
     \end{figure}

\begin{figure}[!ht]
    \centering
       \includegraphics[width=0.49\textwidth]{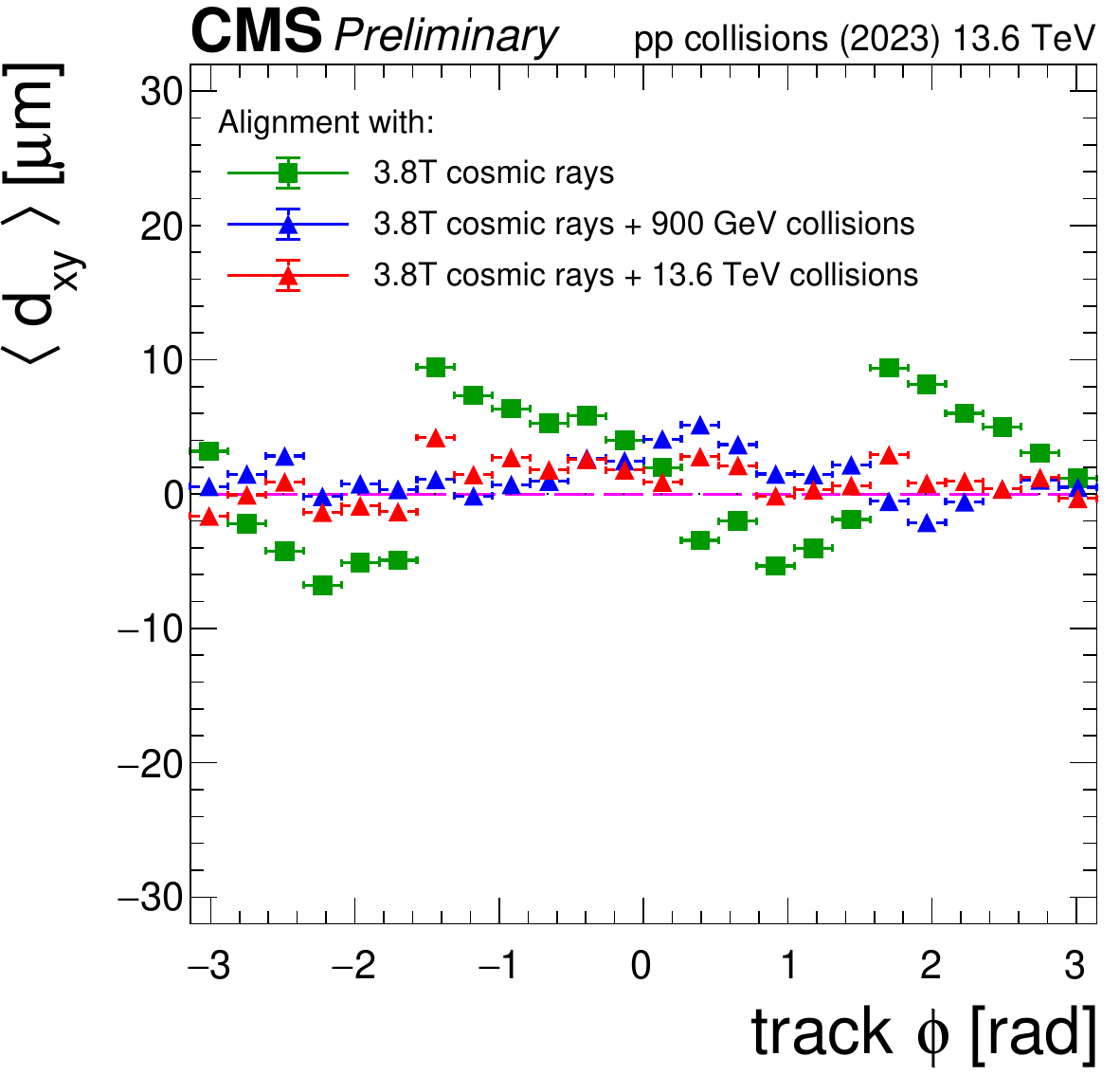}
       \caption{Mean track-vertex impact parameter $d_{xy}$ in bins of track azimuthal angle $\phi$ for three sets of tracker alignment constants derived with cosmic rays and collisions events collected at $\sqrt{s}$=900 GeV and 13 TeV \cite{citation8}.}
       \label{fig2}
     \end{figure}

The pixel detector geometry is monitored by studying the distance between tracks and the vertex, the latter reconstructed without the track under scrutiny (unbiased track-vertex residual) in the transverse plane ($d_{xy}$) and in the longitudinal direction ($d_{z}$), in bins of $\eta$ and $\phi$ of the track. Figure \ref{fig1} (Figure \ref{fig2}) shows RMS of $d_{xy}$ ($d_{xy}$ vs $\phi$) for 2022 (2023) data-taking period.

\subsection{Local reconstruction}
\begin{figure}[!ht]
    \centering
       \includegraphics[width=0.49\textwidth]{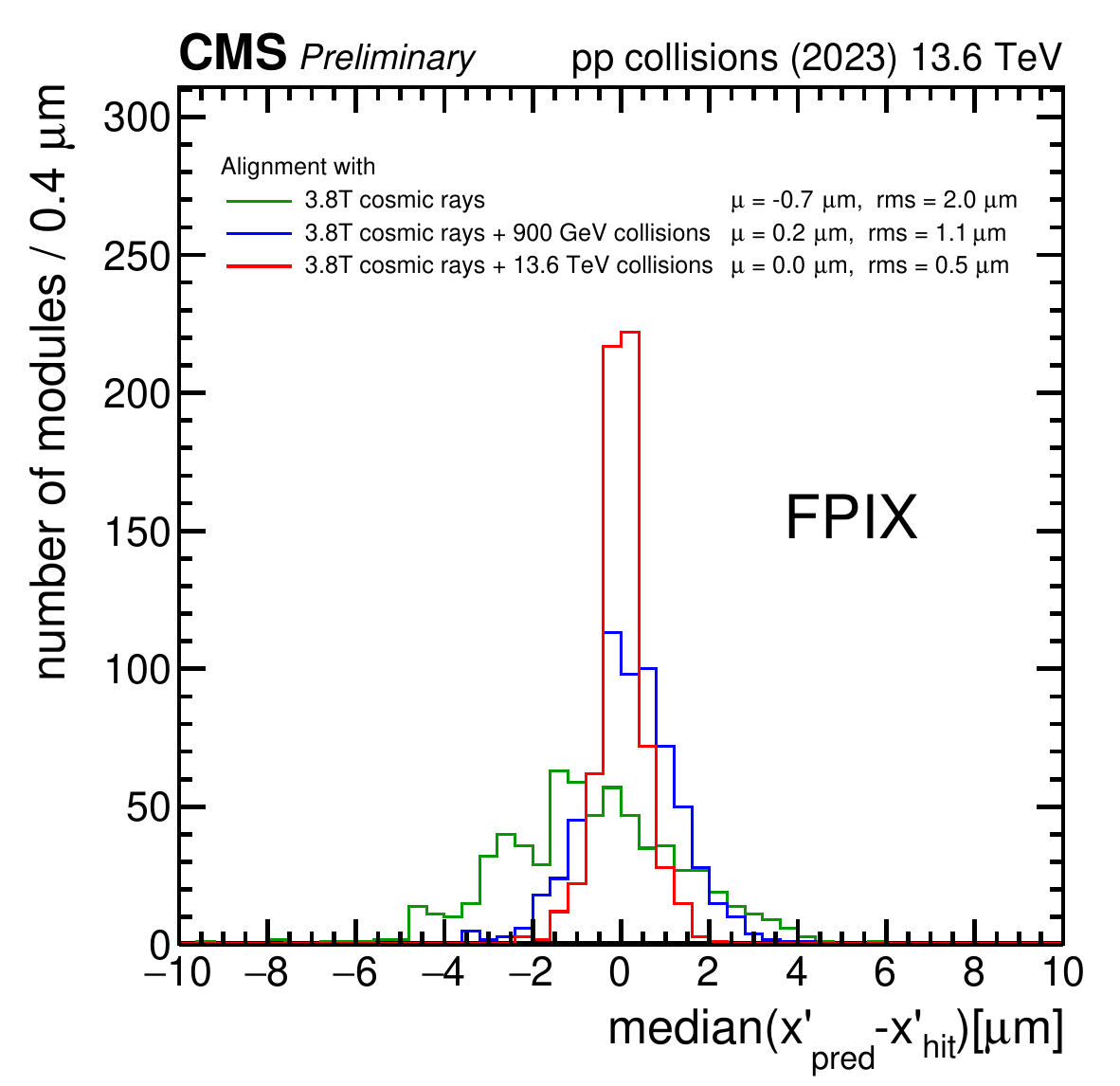}
       \includegraphics[width=0.49\textwidth]{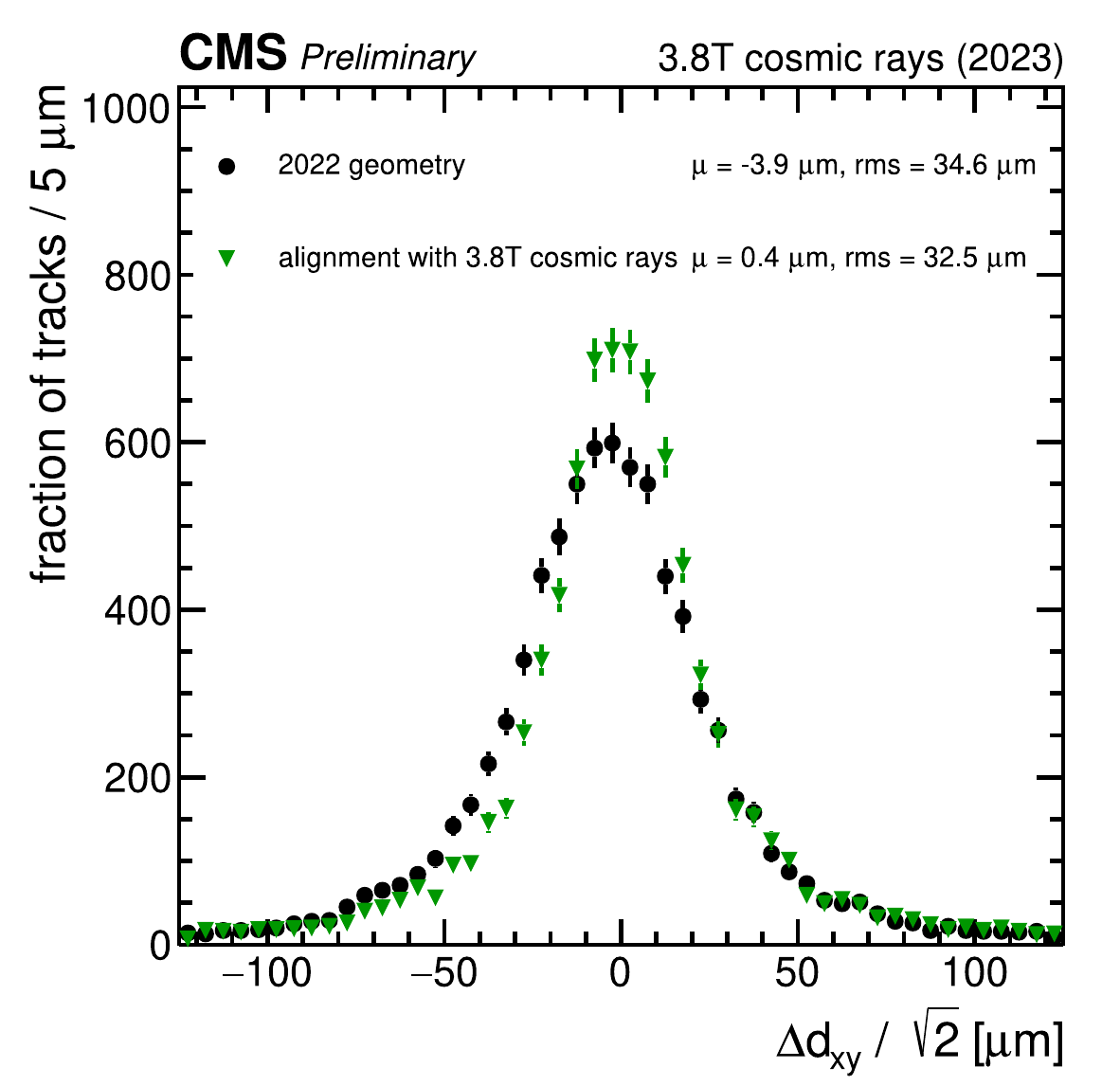}
       \caption{Distribution of median residuals for local-x coordinate in forward pixel detector (left) and difference in track impact parameters in transverse plane $d_{xy}$ (right) between top and bottom halves of cosmic rays \cite{citation8}.}
       \label{fig3}
     \end{figure}

The statistical accuracy of the alignment calibration can be assessed through the distribution of medians of track-hit residuals (DMR), determined with respect to the track prediction, as shown in Figure \ref{fig3} (left), while the track parameter resolution is monitored by independently reconstructing upper and lower portions of cosmic ray muon tracks, and comparing the corresponding track parameters at the point of closest approach to the beamline, as shown in Figure \ref{fig3} (right).
 
\vspace{-0.6cm}

\section{Summary}
The tracker alignment effort during the first two years of LHC Run 3 has been presented. The focus on improving the quality of the alignment calibration already during prompt data reconstruction by optimizing automated workflows has been discussed along with a summary of the excellent Run 3 start in terms of alignment precision as base for derivation of legacy reprocessing conditions.

\end{document}